\documentclass[pre,twocolumn,showpacs,amsmath,amssymb]{revtex4}

\usepackage{graphicx}
\usepackage{dcolumn}
\usepackage{bm}
\usepackage[usenames]{color}

\begin{document}

\preprint{APS/123-QED}

\title{Kinetic models for goods exchange in a multi-agent market}

\author{Carlo Brugna}
\email{carlo.brugna@gmail.com}
\author{Giuseppe Toscani}
\homepage{www-dimat.unipv.it/toscani} \email{giuseppe.toscani@unipv.it}
\affiliation{Dipartimento di Matematica,
  Universit\`a di Pavia,
  Via Ferrata 1,
  27100 Pavia, Italy.}

\date{\today}

\begin{abstract}
 In this paper we introduce a system of kinetic equations describing an exchange market consisting of two populations of agents (dealers and speculators) expressing the same preferences for two goods, but applying different strategies in their exchanges.  Similarly to the model proposed in \cite{TBD}, we describe the trading of the goods by means of some fundamental rules in price theory, in particular by using Cobb-Douglas utility functions for the exchange. The strategy of the speculators is to recover maximal utility from the trade by suitably acting on the percentage of goods which are exchanged.
This microscopic description leads to a system of  linear Boltzmann-type equations for the probability distributions of the goods on the two populations,  in which the post-interaction variables depend from the pre-interaction ones in terms of the mean quantities of the goods present in the market. In this case, it is shown analytically that the strategy of the speculators can drive the price of the two goods towards a zone in which there is a marked utility for their group.  Also, according to \cite{TBD}, the general system of nonlinear kinetic equations of Boltzmann type for the probability distributions of the goods on the two populations is described in details. Numerical experiments then show how the policy of speculators can modify the final price of goods in this nonlinear setting.
\end{abstract}

\maketitle

\def\BigO{ {\mathcal {O}}}
\def\bigM{ {\mathcal {M}}}
\def\bigF{ {\mathcal {F}}}
\def\bigS{ {\mathcal {S}}}
\def\B{\hat \beta}
\def\f{\hat f}
\def\g{\hat g}
\def\Q{\hat Q}
\def\real{\mathbb{R}}
\newcommand{\R}{\mathbb R}
\newcommand{\N}{\mathbb N}
\newcommand{\dt}{\Delta t}
\def\be#1\ee{\begin{equation}#1\end{equation}}
\newcommand{\fer}[1]{(\ref{#1})}
\newtheorem{theorem}{Theorem}[section]
\newtheorem{lemma}[theorem]{Lemma}
\newtheorem{definition}[theorem]{Definition}
\newtheorem{assumption}{Assumption}
\newtheorem{nproposition}[theorem]{Proposition}
\newtheorem{remark}[theorem]{Remark}
\newtheorem{Co}[theorem]{Corollary}
\newtheorem{run}[theorem]{Run}
\newtheorem{example}[theorem]{Example}
\newtheorem{algorithm}{Algorithm}

\setcounter{equation}{0}

\renewcommand{\theequation}{\arabic{section}.\arabic{equation}}
\renewcommand{\thetable}{\arabic{section}.\arabic{table}}
\renewcommand{\thefigure}{\arabic{section}.\arabic{figure}}
\newcommand{\bq}{\begin{equation}}
\newcommand{\eq}{\end{equation}}
\newenvironment{equations}{\equation\aligned}{\endaligned\endequation}
\def\bqa{\begin{eqnarray}}
\def\eqa{\end{eqnarray}}
\def\dt{\Delta t}
\def\m{\mu}
\def\r{\rho}
\def\l{\lambda}
\def\g{\gamma}
\def\e{\epsilon}
\def\a{\alpha}
\def\d{\delta}
\def\t{\theta}
\def\b{\beta}
\def\S{\sigma}
\def\r{\rho}
\def\xt{\tilde x}
\def\yt{\tilde y}


\newcommand{\dis}{\displaystyle}
\newcommand{\bd}{\begin{displaymath}}
\newcommand{\ed}{\end{displaymath}}
\newcommand{\ba}{\begin{eqnarray}}
\newcommand{\ea}{\end{eqnarray}}
\newcommand{\doubleint}{\int\!\!\!\!\int}
\newcommand{\tripleint}{\int\!\!\!\!\int\!\!\!\!\int}
\newcommand{\s}{\sum_{s\,=1}^{4}}
\newcommand{\er}{\sum_{r\,=1}^{4}}
\newcommand{\sr}{\sum_{s,r\,=1}^{4}}
\newcommand{\p}{\partial}
\newcommand{\effe}{{\bf f_{\atop{\,\,}{\!\!\!\!\!\!\!{\tilde{}}}}}}


\def\ff{\widehat f}
\def\lf{{\mathcal L}f}
\def\bb{\hat \beta}
\def\MM{\hat M}
\def\gg{\hat g}
\def\Realr{\mathbb{R}}
\def\N{\mathbb{N}}
\def\R{\mathbb{R}}
\def\var{\varepsilon}
\def\pa{\partial}
\newcommand{\setR}{\mathbb{R}}
\newcommand{\theq}{{\mathbf S}}
\newcommand{\ther}{{\mathbf Q}}
\def\gl{ \tilde g}

\section{Introduction}

In recent years, there has been an increasing interest in developing kinetic models able to describe price formation in a multi-agent society, by resorting to methods typical of statistical mechanics \cite{NPT, PT13}.  In \cite{CPP} Cordier,  Pareschi and Piatecki introduce a kinetic description of the behavior of a simple financial market consisting of a population of agents where each agent can choose to invest between a stock and a bond. In this case, the variation of density is derived starting from the  microscopic model for price formation introduced in
\cite{LLS,LLSb}, usually known as Levy--Levy--Solomon model. The kinetic model proposed in \cite{CPP} attempts to join to simple financial rules a kinetic equation of Boltzmann type, able to describe a complex behavior that could then mimic the market and explain the price formation mechanism.

A further example of coupling wealth with behavioral aspects has been 
proposed in \cite{MD}. This research studies a relatively simple kinetic model for a financial market characterized by a single stock or good and an interplay between two different trader
populations, chartists and fundamentalists, which determine the price dynamics of the stock. The model has been inspired by the microscopic Lux--Marchesi model \cite{LMa,LMb}. The financial rules depends here from the opinion of traders through a kinetic model of opinion formation recently introduced in \cite{To1}. 
A related model has been developed in \cite{BT}, by allowing the opinion variable, which is mainly responsible of the trading, to be strictly connected to price acceleration. 

Also, the importance of the personal knowledge of agents has been recently investigated in  \cite{PT15}, in order to outline how wealth inequality could depend on knowledge distribution in a population.  

In a recent paper, driven by  the assumption that people trades to improve its utility, we coupled  the methods of statistical mechanics and kinetic theory with some principle of price theory in microeconomics \cite{TBD}, considering binary interactions following the rule furnished by the Edgeworth box \cite{Edg}, which is frequently used in general equilibrium theory.
Edgeworth box can fruitfully be applied in presence of an
agent-based system in which agents possess a finite number of goods
of $n\ge 2$ different types. Inspired by this mechanism of
increasing utility and competitive equilibrium,  in \cite{TBD} was introduced and studied
a kinetic equation of Boltzmann type for the evolution
of the distribution density of the quantities of two goods in a system of agents. 
The exchange rule based on the Edgeworth box idea leads to a highly nonlinear binary
interaction, which is difficult to handle, if not numerically. 

For this reason, in \cite{TBD} was considered a suitable linear
Boltzmann equation, obtained by allowing the agent to interact
(according to Edgeworth box), simultaneously with a sufficiently high
number of agents in the market. For this model, it was shown
that this linear equation has a unique solution, and the steady states are concentrated
along a well-defined line (the price line). 

Motivated by the interesting outcomes of the kinetic model based on binary trades driven by the Edgeworth box exchange, and taking into account the intrinsic interest of studying different types of populations  which behave differently with the aim of getting maximum utility \cite{MD, LMa, LMb, PT15}, in what follows we will introduce and discuss a kinetic description of a multi-agent system composed by two populations which interact according to the principle to get maximum utility,  but allowing one of the two populations to exchange goods, by using only a part of them in the cross exchange, with the aim of getting a better profit from this strategy. In analogy with Lux-Marchesi description \cite{LMa,LMb} we will define this population as the population of \emph{speculators}, by leaving the name of \emph{dealers} to the other one.

Applying simple principles of micro--economy \cite{SST}, we first derive in Section \ref{sec-lin}  a linear system of kinetic equations of Boltzmann type, which describes the evolution of the quantities of goods in the two populations. It is shown that the evolution in time of the mean price obeys a non linear law, which in some cases can be explicitly given, to show that the speculators can effectively obtain a net wealth gain from their strategy. 
Then, in Section \ref{nl-model} we will introduce a system of nonlinear kinetic equations, similar to the one considered in \cite{TBD}, which is able do describe the action of a group of speculators in a market of dealers. Numerical experiments, performed in Section \ref{sec-num}, enlighten the possible outcomes of the  various strategies.  

\section{The basic model}\label{model}
As discussed in the introduction, most of the existing kinetic models for
wealth distribution are based on rigid assumptions which, if on one hand
can be shared, from the other hand are not deeply related to economic
principles, like price theory \cite{Fri}. The aim of this Section is to
introduce a framework for trades, which is derived directly from the
basic principles of economy \cite{SST} (cf. also \cite{LLS}). 

Individuals exchange goods. The benefits they receive depend on how
much they exchange and on what terms. Price theory tries to give an
answer to this fundamental question. 

For simplicity, let us start by considering a market with a number $N$ of agents which possess goods of two different types, we denote by $X$ and $Y$.  At the starting time, agents (indexed by $k$) possess certain amounts $x_k= x_k(0)$ of good $X$ and $y_k= y_k(0)$ of good $Y$. While it is clear that $x_k$ and $y_k$ belong to $\N_+$, to avoid inessential difficulties, and without loss of generality, we will always consider these numbers as positive real numbers. The total number of each good in the disposal of agents is given by
 \be\label{tot}
 \sum_{k=1}^N x_k = M_x, \quad  \sum_{k=1}^N y_k = M_y.
 \ee
We assume moreover that the marked is closed, so that the total quantity of goods to be exchanged remains fixed in time. At fixed intervals of time of length $\Delta t$, agents exchange parts of their goods following a certain strategy. In view of these exchanges, agents hold at times $t$ amounts of good $X$ and $Y$, respectively denoted by $x_k(t)$ and $y_k(t)$. By virtue of  \fer{tot}, 
for each time $t\ge\Delta t$
 \be\label{tot1}
 \sum_{k=1}^N x_k(t) = M_x, \quad  \sum_{k=1}^N y_k(t) = M_y.
 \ee
By fixing the price in time of one of the two goods, say $X$, equal to unity, and denoting by $P(t)>0$ the price of the second good $Y$, at each time $t\ge \Delta t$ any agent has a wealth $w_k(t)$ given by
\be\label{wea}
w_k(t) = x_{k}(t_-)+ P(t)  y_{k}(t_-).
\ee 
In \fer{wea}  $t_-= t- \Delta t$ denotes the last exchange time. The same notation will be used in the rest of the paper.
At any subsequent time, agents determine fractions of their wealths, say $x$, to allocate the good $X$, with the reminder allocated to the good $Y$.  In order to give a meaning to the reasons of this trading,  it is classical to
assume that agent's behavior is driven by a utility function.  One of the
most popular of these functions is the {Cobb-Douglas utility function}
 \be\label{CD-0}
 U(x) = x^\alpha (w_k(t) -x)^\beta, \quad \alpha + \beta = 1.
 \ee
Each agent will tend to {maximize its utility} by trading. The values
$\alpha$ and $\beta$ are linked to the preferences that the agent assigns
to the two goods. If $\alpha >\beta$, the agent prefers to possess goods
of the first type (numbered by $x$). The choice $\alpha= \beta = 1/2$
clearly means that the two goods are equally important for $A$. The maximization of \fer{CD-0} updates the quantities of goods to
 \be\label{max3}
 x_k(t)= \alpha w_k(t); \quad P(t)  y_{k}(t) = \beta w_k(t).
 \ee
The value of the unknown price $P(t)$ can be easily determined from the values of the variables at time $t-\Delta t$ by resorting to the constraint \fer{tot1}. Indeed it holds
 \be\label{const}
  \sum_{k=1}^N \alpha w_k(t) = M_x; \quad  \sum_{k=1}^N \beta w_k(t) = P_k(t)M_y.
 \ee
Hence, solving for $P(t)$ we obtain the (fixed in time) price
 \be\label{newp}
 P(t) = P= \frac{ \beta M_x}{\alpha M_y}.
 \ee
As expected, the (relative) price \fer{newp} of the good $Y$  is directly proportional to its preference value $\beta$, and inversely proportional to the quantity $M_y$ of goods of type $Y$ present in the closed marked. 
Substituting the expression for $P(t)$ back into \fer{max3} leads to the new quantities of goods
\begin{equations}\label{n5}
& x_k(t) = x_k(t_-) + \beta \left( \frac{M_x}{M_y} y_k(t_-)- x_k(t_-)\right) \\
& y_k(t) = y_k(t_-) + \alpha\left( \frac{M_y}{M_x} x_k(t_-)- y_k(t_-)\right).
\end{equations}
This very simple mechanism of exchange can be easily generalized to different groups of traders, which can adopt different strategies. The simplest of these generalizations is to consider a market composed by two groups of agents, say $A$ and $B$, with  fixed numbers $N_A$  and $N_B$ of agents belonging to the groups $A$ and $B$. As in the previous case,  agents of both groups possess goods of two different types $X$ and $Y$.  At the starting time, agents of the group $A$ possess certain amounts $x_k= x_k(0)$ of good $X$ and $y_k= y_k(0)$ of good $Y$, and  the total number of each good in the disposal of agents of group $A$ is given by
 \be\label{totA}
 \sum_{k=1}^{N_A} x_k = M_x, \quad  \sum_{k=1}^{N_A} y_k = M_y.
 \ee
Likewise, agents of the group $B$ possess certain amounts $\xt_k= \xt_k(0)$ of good $X$ and $\yt_k= \yt_k(0)$ of good $Y$, and  the total number of each good in the disposal of agents of group $B$ is given by
 \be\label{totB}
 \sum_{k=1}^{N_B} \tilde x_k = m_x, \quad  \sum_{k=1}^{N_B} \tilde y_k = m_y.
 \ee
By assuming that the marked is closed, the total quantity of goods at disposal of the two groups is conserved in time. However, since agents of the two groups interact on the same market, the total number of  goods in each population can change with time. Therefore at time $t \ge \Delta t$
\be\label{totA-t}
 \sum_{k=1}^{N_A} x_k(t) = M_x(t), \quad  \sum_{k=1}^{N_A} y_k(t) = M_y(t), 
 \ee
and
\be\label{totB-t}
 \sum_{k=1}^{N_B} \tilde x_k(t) = m_x(t), \quad  \sum_{k=1}^{N_B} \tilde y_k = m_y(t).
 \ee
The conservation of goods of type $X$ and $Y$ then implies
 \be\label{con5}
 M_x(t) + m_x(t) = M_x + m_x, \quad M_y(t) + m_y(t) = M_y + m_y
 \ee 
at any subsequent time $t\ge\Delta t$. 
At integer times $t\ge\Delta t$, agents of  exchange parts of their goods. Agents of group $A$, the dealers,  follows the previous strategy. Agents of group $B$, the speculators, follows a different strategy, essentially based on saving. While possessing amounts $(\xt_k(t_-), \yt_k(t_-))$ of goods, they exchange on the market only a  part $(\lambda_x \xt_k(t_-), \lambda_y \yt_k(t_-))$ of their goods, where $0 < \lambda_x, \lambda_y < 1$ are fixed constants. 
Consequently, the wealth of the $k$-th speculator, say $\tilde w_k(t)$ involved in the exchange is
\be\label{wea-c}
\tilde w_k(t) = \lambda_x x_{k}(t_-)+ P(t) \lambda_y y_{k}(t_-).
\ee 
Hence, in presence of this saving policy, the total  quantity of goods to be exchanged does not coincide with the total (fixed) number of goods available in the market, given by the sum of the quantities \fer{totA} and \fer{totB}. The saving policy of speculators introduces the (time dependent) constraints
 \begin{equations}\label{n-con}
 & \sum_{k=1}^{N_A} \alpha w_k(t)+  \sum_{k=1}^{N_B} \alpha \tilde w_k(t) = M_x(t_-) + \lambda_xm_x(t_-), \\
 &\sum_{k=1}^{N_A} \beta w_k(t)+  \sum_{k=1}^{N_B} \beta \tilde w_k(t) = P(t)(M_y(t_-) + \lambda_y m_y(t_-)),
 \end{equations}
that express the effective quantities of goods present in the market at time $t-1$. Proceeding as before, agents of group $A$ update their goods as in \fer{max3}. However, using the new constraints \fer{n-con} gives the time-dependent price
 \be\label{n-pri}
P(t) = \frac{ \beta(M_x(t_-) + \lambda_xm_x(t_-)) }{\alpha (M_y(t_-) + \lambda_y m_y(t_-))}. 
 \ee
Substituting the expression for $P(t)$ back into \fer{wea} leads for dealers to the new quantities of goods
\begin{equations}\label{n6}
 x_k(t) = &x_k(t_-) + 
 \\
 +&\beta \left(\frac{M_x(t_-) + \lambda_xm_x(t_-) }{M_y(t_-) + \lambda_y m_y(t_-)} y_k(t_-)- x_k(t_-)\right) \\
y_k(t) = &y_k(t_-) +
\\ 
+& \alpha\left(\frac{M_y(t_-) + \lambda_y m_y(t_-)}{M_x(t_-) + \lambda_xm_x(t_-) }  x_k(t_-)- y_k(t_-)\right).
\end{equations}
Likewise, speculators update their quantities of goods by
\begin{equations}\label{n7}
\xt_k(t) =& \xt_k(t_-) + 
\\
+&\beta \left(\frac{M_x(t_-) + \lambda_xm_x(t_-) }{M_y(t_-) + \lambda_y m_y(t_-)} \lambda_y \yt_k(t_-)- \lambda_x \xt_k(t_-)\right) \\
\yt_k(t) = &\yt_k(t_-) +
\\
+& \alpha\left(\frac{M_y(t_-) + \lambda_y m_y(t_-)}{M_x(t_-) + \lambda_xm_x(t_-) }\lambda_x  \xt_k(t_-)- \lambda_y \yt_k(t_-)\right).
\end{equations}
Similar expressions for the updated quantities of goods have been obtained in \cite{TBD} by resorting to the binary exchange rule provided by the Hedgeworth box, and subsequently linearizing the outcome. We will be back to this analogy later on.

\section{A  Boltzmann system for trading of goods}\label{sec-lin}
\subsection{The kinetic model}
The previous model will be now modelled within the principles of statistical mechanics. Let the multi-agent system under study be composed by the two classes of agents of section \ref{model}.   Let  $f(x,y,t)$ denote the density of agents of the class A (the dealers) with quantities $x$ and $y$
of the two goods at time $t \geq 0$, and let $g(x,y,t)$ denote the density of agents of the class $B$ (the speculators) with quantities $x$ and $y$ of the two goods at time $t \geq 0$. As before, and without loss of generality, we will assume  that $x$ and $y$ are nonnegative real numbers.  A system of Boltzmann-like equations of Maxwell type for the time evolution of the two densities
$f(x,y,t)$ and $g(x,y,t)$ can be written  in terms of the updated quantities \fer{n6} and \fer{n7} according to the following assumptions. The total number of each good in the disposal of agents at time $t \ge 0$, previously given by \fer{totA-t} and \fer{totB-t} is here substituted by the mean values  of the density functions at time $t \ge 0$
 \begin{equations}\label{ftotA}
 &M_x(t)= \int_{\R_+} x\, f(x,y, t)\, dx\, dy,
 \\
 &  M_y(t)= \int_{\R_+} y\, f(x,y, t)\, dx\, dy,
 \end{equations}
and
 \begin{equations}\label{ftotB}
 &m_x(t)= \int_{\R_+} x\, g(x,y, t)\, dx\, dy, 
 \\
 & m_y(t)= \int_{\R_+} y\, g(x,y, t)\, dx\, dy.
 \end{equations}
We remark that this weaker assumption on the amount of goods effectively present in the market appears realistic, and it is consistent with the fact that agents in the market have an exact perception of the quantities of goods only in terms of their mean values.  
Moreover, to take into account that it is very difficult to exchange goods by performing an optimal exchange, we allow the updating of goods to be dependent of some randomness expressing deviation from the optimal choice, in any case by maintaining optimality in mean sense. In agreement with \cite{TBD} this can be done by taking the values $0\le \alpha(\omega)\le 1$ and $0 \le \beta(\omega)\le 1$ in the updated quantities \fer{n6} and \fer{n7} as positive independent random variables  such that
 \be\label{con7}
 \langle \alpha(\omega) \rangle = \alpha, \quad \langle \beta(\omega) \rangle = \beta, \quad \alpha + \beta = 1,
 \ee 
where $\langle \cdot\rangle$ denotes mathematical expectation.

In reason of these choices, given a quantity $(x,y)$ of goods at time $t \ge 0$, the dealers will update their quantities according to
\begin{equations}\label{trA}
  &x^* = x + \beta(\omega) \left( \frac{M_x(t) +\lambda_x m_x(t)}{M_y(t) +\lambda_y m_y(t)} y - x\right) \\
   & y^* = y + \alpha(\omega) \left( \frac{M_y(t) +\lambda_y m_y(t)}{M_x(t) +\lambda_x m_x(t)} x - y\right).
   \end{equations}
   Likewise, the speculators will update at time $t \ge 0$  the quantities $(x, y)$ according to
   \begin{equations}\label{trB}
  &\tilde x^* = x + \beta(\omega) \left( \frac{M_x(t) +\lambda_x m_x(t)}{M_y(t) +\lambda_y m_y(t)} \lambda_y \, y - \lambda_x \, x\right) \\
   & \tilde y^* =  y + \alpha(\omega) \left( \frac{M_y(t) +\lambda_y m_y(t)}{M_x(t) +\lambda_x m_x(t)} \lambda_x \, x -\lambda_y \, y\right).
   \end{equations}
Note that the law of variation of the goods in disposal at time $t>0$ depends on the mean values of the densities of the two groups at the same time. 

In agreement with definition \fer{n-pri}, the mean price $P(t)$ at time $t \ge 0$ of the second good relative to the first one is defined by
 \be\label{newpp}
 P(t) = \frac {\beta}{\alpha} \frac{M_x(t) +\lambda_x m_x(t)}{M_y(t) +\lambda_y m_y(t)}.
 \ee
It is immediate to recognize that both interactions of type \fer{trA} and \fer{trB} imply the  conservation in the mean, at each time $t \ge 0$, of the agent's wealths. Indeed
 \begin{equations}\label{con-w}
w_A^*(t)=& x^* + P(t)y^* = 
\\
& x+ P(t)y + \left(\beta(\omega) - \frac \beta\alpha \alpha(\omega)\right)\left(\frac\alpha\beta P(t) y - x\right), \\
\tilde w_B^*(t)= &\tilde x^* + P(t)\tilde y^* =\tilde x+ P(t)\tilde y \,+ 
\\
+ & \left(\beta(\omega) - \frac \beta\alpha \alpha(\omega)\right)\left(\frac\alpha\beta P(t)\lambda_y y - \lambda_x x\right),
 \end{equations}
which implies
\begin{equations}\label{con-ww}
&\langle w_A^*(t)\rangle = x+ P(t)y = w_A(t), 
\\
&  \langle\tilde w_B^*(t)\rangle =  x+ P(t) y =w_B(t).
\end{equations} 
Once the mechanism of variation of the quantities of goods has been defined, the evolution in time of the densities can be easily written down by expressing the law of variation in time of the observable quantities.  It corresponds to write, for any given smooth function
$\varphi$, a system of linear spatially homogeneous Boltzmann equations
 \begin{equations}\label{we}
&\frac{d}{dt} \int_{\real_+^2}  \varphi(x, y) f(x,y,t)\,dx \, dy =
\\
& \sigma \left\langle \int_{\real_+^2}  ( \varphi(x^* ,y^*)-\varphi(x
,y))f(x,y,t) \,dx \, dy  \,\right\rangle , \\
&\frac{d}{dt} \int_{\real_+^2}  \varphi(x, y) g(x,y,t)\,dx \, dy = 
\\
&\sigma \left\langle \int_{\real_+^2}  ( \varphi(\tilde x^* ,\tilde y^*)-\varphi(x
,y))g(x,y,t) \,dx \, dy  \,\right\rangle.
\end{equations}
In \fer{we} the positive constant $\sigma$ is a measure of the frequency of interactions.  The right-hand sides of equations \fer{we} describe the change of
 density $f$ due the variation of type \fer{trA} (respectively the change of $g$  due the variation of type \fer{trB}). The two kinetic equations in \fer{we} are linked each other in view of the nature of the microscopic interactions, which involve the mean values of both densities.
By choosing $\varphi(x, y) = x$ (respectively $\varphi(x, y) = y$)
one reckons the laws of variation in time of the mean values \fer{ftotA} and \fer{ftotB} relative to the populations of dealers and speculators.  The mean values of the population of dealers change according to
\begin{equations}\label{meA}
& \frac {dM_x(t)}{dt} = \beta \left( \frac{M_x(t) +\lambda_x m_x(t)}{M_y(t) +\lambda_y m_y(t)} M_y(t) - M_x(t)\right), \\
& \frac {dM_y(t)}{dt} = \alpha \left( \frac{M_y(t) +\lambda_y m_y(t)}{M_x(t) +\lambda_x m_x(t)} M_x(t) - M_y(t)\right). \end{equations}
Likewise, the mean values of the population of speculators change according to 
\begin{equations}\label{meB}
& \frac {dm_x(t)}{dt} = \beta \left( \frac{M_x(t) +\lambda_x m_x(t)}{M_y(t) +\lambda_y m_y(t)} \lambda_y m_y(t) - \lambda_x m_x(t)\right), \\
& \frac {dm_y(t)}{dt} = \alpha \left( \frac{M_y(t) +\lambda_y m_y(t)}{M_x(t) +\lambda_x m_x(t)} \lambda_x m_x(t) - \lambda_y m_y(t)\right). 
\end{equations}
It is immediate to recognize that 
\begin{equations}\label{conXY}
& M_x(t) + m_x(t) = I_x= M_x(0)+ m_x(0), \\
& M_y(t) + m_y(t) = I_y= M_y(0)+ m_y(0).
\end{equations}
These constraints reflect the conservation of the mean quantities of goods present in the closed market. Also, in view of \fer{con-ww} it holds
\begin{equations}\label{evoW}
& \frac d{dt}W_A(t) =  M_y(t)\frac d{dt}P(t), \\
& \frac d{dt}W_B(t) =  m_y(t)\frac d{dt}P(t).
\end{equations}
Equations \fer{evoW} express the time variation of the mean wealths of the agents of classes $A$ and $B$, in terms of the evolution of the price. 
\begin{remark} It is interesting to remark that, provided that both the saving constants $\lambda_x, \lambda_y$ are strictly positive, we have a further conservation law. 
Thanks to \fer{conXY}
 \begin{equations}\label{equ8}
& M_x(t) +\lambda_x m_x(t) = I_x - (1-\lambda_x)m_x(t),\\
&  M_y(t) +\lambda_y m_y(t) =I_y-  (1-\lambda_y)m_y(t).  
 \end{equations}
A further conservation follows considering that, if for any given $r >0$
\be\label{log-x}
\Phi_x (r) = \frac1{\beta(1-\lambda_x)} \log{\left[ I_x - (1-\lambda_x)r \right]},
\ee
it holds
\begin{equations}\label{der4}
&- \frac{d\Phi_x (m_x(t))}{dt} = - \Phi_x'(m_x(t)) \frac {dm_x(t)}{dt} =  \\
& + \frac 1{\beta\left[I_x - (1-\lambda_x)m_x(t) \right]}\frac {dm_x(t)}{dt} = \\
&\frac {\lambda_y m_y(t)} {I_y - (1-\lambda_y)m_y(t)} - \frac {\lambda_x m_x(t)} {I_x - (1-\lambda_x)m_x(t)} .
\end{equations}
In analogous way, if 
\be\label{log-y}
\Phi_y(r) = \frac1{\alpha(1-\lambda_y)} \log{ \left[I_y - (1-\lambda_y)r \right]},
\ee
it holds
\begin{equation}\label{der5}
- \frac{d\Phi_y(m_y(t))}{dt} =  \frac{\lambda_x m_x(t) }{I_x - (1-\lambda_x)m_x(t) } - \frac {\lambda_y m_y(t)} {I_y - (1-\lambda_y)m_y(t)} \end{equation}
Consequently
\[
\frac{d\left[ \Phi_x (m_x(t)) + \Phi_y(m_y(t))\right]}{dt} = 0,
\]
which implies the conservation law
 \begin{equations}\label{con6}
& \left(I_x-  (1-\lambda_x)m_x(t) \right)^{1/[\beta(1-\lambda_x)]}\cdot \\
&\cdot\left( I_y- (1-\lambda_y)m_y(t) \right)^{1/[\alpha(1-\lambda_y)]} = C_0,
  \end{equations}
\end{remark}
Since the laws of evolution \fer{meA} and \fer{meB} are nonlinear, even in presence of conservation laws, a precise analytical study of systems \fer{meA} and \fer{meB} appears very  difficult. Likewise, it is cumbersome to find the evolution of the higher moments of the solutions to \fer{we} in a closed form. Hence, the Boltmann equation \fer{we} is the starting point for a numerical study of the evolution of the densities by means of Monte Carlo methods \cite{BT,Pa}.

\subsection{An explicitly solvable case}
In what follows, we will discuss the situation in which the agents of the class $B$ will trade on the closed market only goods of one type, say $Y$, with the intent to increase their quantity of goods of type $X$. This corresponds to choose $\lambda_x=0$, and $\lambda_y = \delta$. The simplest case is obtained when $\lambda_y=1$, namely the case in which agents of the second class will exchange into the marked the total amount of their goods of type $Y$. 

If this is the case, the second equation in \fer{meB} reduces to
 \be\label{vab}
\frac {dm_y(t)}{dt} = - \alpha m_y(t), 
\ee 
which can be easily solved to give
 \be\label{me-y} 
 m_y(t) = m_y(0) \exp \{ -\alpha t\}.
 \ee
 This shows that the goods of type $Y$ in the hands of the class $B$ of agents is exponentially decreasing in time at a rate proportional to $\alpha$, and the class $B$ will remain only with goods of type $Y$. 
 
 Then,  owing to the conservations \fer{conXY} the first equation in \fer{meA} takes the form
  \begin{equations} \label{me-x}
 \frac {dM_x(t)}{dt} = &\beta \left( \frac{M_x(t)}{M_y(t) + m_y(t)} M_y(t) - M_x(t)\right) =
 \\
- & \beta M_x(t) \frac{m_y(t)}{I_y}.  \end{equations}
 Hence we have
   \be\label{me-x2}
 \frac {d\log M_x(t)}{dt} =  -\beta \frac{m_y(0)}{I_y} \exp\{-\alpha t\},
   \ee 
 that can be integrated to give
  \be\label{M_x}
  M_x(t) = M_x(0) \exp\left\{ - \frac \beta\alpha \frac{m_y(0)}{I_y}\left( 1-  \exp\{-\alpha t\}\right) \right\} .
  \ee
 As expected, the quantity of goods of type $X$ in the class $A$ of agents is decreasing in time, and will exponentially reach the limit value
  \[
 \bar M_x = M_x(0) \exp\left\{ - \frac \beta\alpha \frac{m_y(0)}{I_y} \right\} .
    \] 
 Owing to the conservation laws \fer{conXY},  from  \fer{me-y} and \fer{M_x} we then obtain the values of $m_x(t)$ and $M_y(t)$. Also, the mean price $P(t)$ defined in \fer{newpp} is
 \be\label{pri1}
 P(t) = \frac\beta\alpha \frac{M_x(0)}{I_y} \exp\left\{ - \frac \beta\alpha \frac{m_y(0)}{I_y}\left( 1-  \exp\{-\alpha t\}\right) \right\} .
  \ee
 The price of the good $Y$ relative to $X$ is decreasing in time, and it will reach the limit value
  \[
  \bar P =  \frac\beta\alpha \frac{M_x(0)}{I_y} \exp\left\{ - \frac \beta\alpha \frac{m_y(0)}{I_y} \right\} .
  \]
 Note that in this case the time variation of the wealths  of the two classes, given by  \fer{evoW} is negative, and both classes will show their wealths decrease in time. 
 
 Consider the case in which initially the two classes possess the same mean quantity of the good $Y$, but the class $A$ possesses a bigger mean quantity of the good $X$. Then, at time $t=0$ one has $W_A(0) > W_B(0)$. Since the class $B$ is transferring goods of type $Y$ to the class $A$,  at any subsequent time $t >0$ the difference $M_y(t) - m_y(t)$ is positive, and  the relative wealth $W_A(t) - W_B(t)$, in reason of \fer{evoW} is decreasing in time.
 In this case, the strategy of the agents of the class $B$ is such that its final wealth is closer to the wealth of the class $A$.
 
 The previous example shows that the eventual strategy of the agents of the class $B$ is able to determine an effective improvement of their wealth conditions. 

\subsection{The general case}

Let us set 
\begin{equations}\label{def-rho}
& \rho_x(t) = \frac{M_x(t)}{M_x(t) + \lambda_x m_x(t)},\\
& \rho_y(t) = \frac{M_y(t)}{M_y(t) + \lambda_y m_y(t)}.
\end{equations}
 Then, equations \fer{meA} can be rewritten as
 \begin{equations}\label{meAA}
&\frac{1}{M_x(t) + \lambda_x m_x(t)} \frac {dM_x(t)}{dt} = \beta \left(\rho_y(t) - \rho_x(t)\right), \\
& \frac{1}{M_y(t) + \lambda_y m_y(t)} \frac {dM_y(t)}{dt} = \alpha \left(\rho_x(t) - \rho_y(t)\right). \end{equations}
Using the constraints \fer{conXY} it follows that
 \be\label{der-x}
  \frac {d\rho_x(t)}{dt} = \frac{\lambda_x I_x}{\left( \lambda_xI_x + (1-\lambda_x)M_x(t)\right)^2}\frac {dM_x(t)}{dt},
 \ee
 and analogous result (changing $x$ with $y$) for $\rho_y(t)$. Hence, since \fer{def-rho} imply
 \be\label{rel-rho}
 \frac{\lambda_x I_x}{\lambda_x I_x + (1-\lambda_x)M_x(t)} = 1- (1-\lambda_x)\rho_x,
 \ee
equations \fer{meAA}  take the form
  \begin{equations}\label{meA-A}
& \frac {d\rho_x(t)}{dt} = \beta (1- (1-\lambda_x)\rho_x(t)) \left(\rho_y(t) - \rho_x(t)\right), \\
&  \frac {d\rho_y(t)}{dt} = \alpha (1- (1-\lambda_y)\rho_y(t))\left(\rho_x(t) - \rho_y(t)\right). \end{equations}
 It is clear that the equilibrium points of system \fer{meA-A} are obtained when $\rho_x= \rho_y$. on the other hand, the conservation law \fer{con6} implies that, in equilibrium the function
 \begin{equations}\label{con8}
 H(\rho_x(t), \rho_y(t))= & \left( (1- (1-\lambda_x)\rho_x(t))\right)^{1/\beta(1-\lambda_x)}\cdot \\
 & \left( (1- (1-\lambda_y)\rho_x(t))\right)^{1/\alpha(1-\lambda_x)} 
 \end{equations}
 satisfies the identity
 \be\label{id3}
  H(\rho_x, \rho_y)= H(\rho_x, \rho_x)= H(\bar\rho_x, \bar\rho_y),
 \ee
where $\bar\rho_x, \bar\rho_y$ denote the initial values. Since the function $H=H(u,v)$ is decreasing  with respect to both $u$ and $v$ when $0<u,v<1$, whenever $\bar\rho = \min\{\bar\rho_x, \bar\rho_y \}$
 \[
 H(\bar\rho_x, \bar\rho_y)\ge H(\bar\rho, \bar\rho).
 \]
Hence, there is a unique equilibrium point $\rho$ with $\min\{\bar\rho_x, \bar\rho_y \}\le \rho\le \max\{\bar\rho_x, \bar\rho_y \}$ in which 
 \[
 H(\rho, \rho)= H(\bar\rho_x, \bar\rho_y).
 \]
 In correspondence to this equilibrium point, we obtain the limit value of the price
 \be\label{p-lim}
 \bar P = \frac{\beta \lambda_ x I_x}{\alpha \lambda_ y I_y}\frac{1- (1-\lambda_y)\rho}{1- (1-\lambda_x)\rho}.
 \ee

 \section{A system of nonlinear Boltzmann equations}\label{nl-model}
The basic model discussed in Section \ref{sec-lin} can be easily adapted to recover a microscopic binary interaction between agents of the same class, or between agents of different classes. 

For simplicity, let us start as in Section \ref{model} by considering a market with a number $N$ of agents which possess goods of two different types $X$ and $Y$.  Consider now at time $t$ a trading between two agents with wealths $w_j(t)$ and $w_k(t)$ given respectively  by
\begin{equations}\label{we-jk}
&w_j(t) = x_{j}(t_-)+ P(t)  y_{j}(t_-), 
\\
& w_k(t) = x_{k}(t_-)+ P(t)  y_{k}(t_-).
\end{equations} 
Then, both agents   determine a fractions of their wealths to allocate the good $X$, with the reminder allocated to the good $Y$.  Using again the {Cobb-Douglas utility function} \fer{CD-0}
agents will update the quantities of goods to
 \begin{equations}\label{max-jk}
 &x_j(t)= \alpha w_j(t), \quad P(t)  y_{j}(t) = \beta w_j(t), 
 \\
 & x_k(t)= \alpha w_k(t), \quad P(t)  y_{k}(t) = \beta w_k(t).
  \end{equations}
At difference with the previous (global) analysis of the closed market, the value of the  price $P(t)$ is now determined from the values of the variables at time $t-\Delta t$ by resorting to the constraints of  conservation of the sums  of goods of type $X$ and $Y$ of the two agents $j$ and $k$. In this case
  \begin{equations}\label{con-jk}
  & \alpha(w_j(t)+ w_k(t) ) = x_j(t_-)+x_k(t_-), \\
  &   \beta (w_j(t) + w_k(t) )= P_k(t)(y_j(t_-) + y_k(t_-)).
  \end{equations}
Hence, solving for $P(t)$ we obtain the relation
 \be\label{ppp}
 P(t) =  \frac{ \beta }{\alpha}\frac{ x_j(t_-)+x_k(t_-)}{y_j(t_-) + y_k(t_-)}.
 \ee
Substituting the expression for $P(t)$ back into \fer{max3} leads for the agent $j$ to the new quantities of goods
\begin{equations}\label{n55}
& x_j(t) = x_j(t_-) + \beta \left( \frac{x_j(t_-)+x_k(t_-)}{y_j(t_-) + y_k(t_-)} y_j(t_-)- x_j(t_-)\right) \\
& y_j(t) = y_j(t_-) + \alpha\left(   \frac{y_j(t_-) + y_k(t_-)}{x_j(t_-)+x_k(t_-)} x_j(t_-)- y_j(t_-)\right).
\end{equations}
Analogous result holds for the updating of the quantities of goods of the agent $k$.

This binary  exchange can be easily generalized to the two groups of dealers and speculators considered before. In a binary exchange between the dealer $j$ of the class $A$ and the speculator $k$  of the class $B$, the dealer $j$  updates its goods according to
\begin{equations}\label{n65}
& x_j(t) = x_j(t_-) + \beta \left(\frac{x_j(t_-) + \lambda_x x_k(t_-) }{y_j(t_-) + \lambda_y y_k(t_-)} y_j(t_-)- x_j(t_-)\right), \\
& y_j(t) = y_j(t_-) + \alpha\left(\frac{y_j(t_-) + \lambda_y y_k(t_-)}{x_j(t_-) + \lambda_x x_k(t_-) }  x_j(t_-)- y_j(t_-)\right).
\end{equations}
Likewise, a speculator $k$  updates its quantities of goods according to
\begin{equations}\label{n66}
 x_k(t) =& x_k(t_-) +
 \\
  +&\beta \left(\frac{x_j(t_-) + \lambda_x x_k(t_-) }{y_j(t_-) + \lambda_y y_k(t_-)} \lambda_y y_k(t_-)- \lambda_x x_k(t_-)\right) ,\\
 y_k(t) = &y_k(t_-) + 
 \\
 +& \alpha\left(\frac{y_j(t_-) + \lambda_y y_k(t_-)}{x_j(t_-) + \lambda_x x_k(t_-) }  \lambda_k x_k(t_-)- \lambda_y y_k(t_-)\right).
\end{equations} 
We remark that similar expressions for the updated quantities of goods have been obtained in \cite{TBD} by resorting to the binary exchange rule provided by the Hedgeworth box.  Similarly to the model considered of \cite{TBD} we can use the binary exchanges \fer{n55} , \fer{n65} and \fer{n66} to construct a system of bilinear Boltzmann type equations. 

We assume that the dealers can interact with other dealers, and with speculators, according to their respective rules \fer{n55} and \fer{n65}, while speculators interact only with dealers, according to \fer{n66}.  As before, let  $f(x,y,t)$ denote the density of dealers of the class  $A$ with quantities $x$ and $y$
of the two goods at time $t \geq 0$, and let $g(x,y,t)$ denote the density of speculators of the class $B$ with quantities $x$ and $y$ of the two goods at time $t \geq 0$. Then $f(x,y,t)$ satisfies
 \begin{equations}\label{we-f}
\frac{d}{dt} \int_{\real_+^2}  \varphi(x, y)& f(x,y,t)\,dx \, dy = \\
&\sigma \int_{\real_+^2}\varphi(x ,y)Q(f,f)(x,y) \,dx dy + \\
+ &\mu \int_{\real_+^2}\varphi(x ,y)P(f,g)(x,y) \,dx dy  .
 \end{equations}
The constants $\sigma$ and $\nu$ measure the frequency of collisions. The right-hand side of equation \fer{we} describes the change of density due to exchanges between dealers (the operator $Q$) and exchanges dealer--speculator (the operator $P$). The definitions of $Q$ and $P$ are fruitfully given by their action on observable quantities
\begin{equations}\label{Q1}
& \int_{\real_+^2}\varphi(x ,y)Q(f,f)(x,y) \,dx  dy =\\
&\left\langle \int_{\real_+^4}  ( \varphi(x^* ,y^*)-\varphi(x
,y))f(x,y,t)f(x_1,y_1,t)  \, d\pi \right\rangle,
 \end{equations}
 where, now and later on,  $d\pi = dx \, dy \, dx_1 \, dy_1$, and
 \begin{equations}\label{AA}
  &x^* = x + { \beta(\omega)} \left( \frac{x +x_1}{y + y_1} y - x\right), \\
   & y^* = y + {\alpha(\omega)} \left( \frac{y +y_1}{x + x_1} x -
   y\right).
   \end{equations}
   Likewise 
\begin{equations}\label{P1}
& \int_{\real_+^2}\varphi(x ,y)P(f,g)(x,y) \,dx  dy =
\\
& \left\langle \int_{\real_+^4} \,
( \varphi(\tilde x ,\tilde y)-\varphi(x
,y))f(x,y,t)g(x_1,y_1,t)  \, d\pi\right\rangle,
 \end{equations}
  where
\begin{equations}\label{AB}
  &\tilde x = x + { \beta(\omega)} \left( \frac{x +\lambda_x x_1}{y + \lambda_y y_1} y - x\right), \\
   &\tilde y = y + {\alpha(\omega)} \left( \frac{y +\lambda_y y_1}{x + \lambda_x x_1} x -
   y\right).
   \end{equations} 
In agreement with Section \ref{sec-lin} the values $0\le \alpha(\omega)\le 1$ and $0 \le \beta(\omega)\le 1$ in  \fer{AA} and \fer{AB} are positive independent random variables satisfying
 \fer{con7}.   
Equation \fer{we-f} is coupled with the evolution equation for $g(x,y,t)$, given by   
 \begin{equations}\label{we-g}
\frac{d}{dt} \int_{\real_+^2}  \varphi(x, y)& f(x,y,t)\,dx \, dy = 
\\
& \mu \int_{\real_+^2}\varphi(x ,y)\bar P(g,f)(x,y) \,dx dy .
 \end{equations}
 Clearly
\begin{equations}\label{P2}
& \int_{\real_+^2}\varphi(x ,y)\bar P(g,f)(x,y) \,dx  dy =
\left\langle \int_{\real_+^4} \,dx \, dy \, dx_1 \, dy_1 \right.
\\
&\left.  ( \varphi(\bar x ,\bar y)-\varphi(x
,y))g(x,y,t)f(x_1,y_1,t)\,\right\rangle,
 \end{equations}
  where
\begin{equations}\label{AB1}
  &\bar x = x + { \beta(\omega)} \left( \frac{x_1 +\lambda_x x}{y_1 + \lambda_y y}\lambda_y y - \lambda_x x\right), \\
   &\tilde y = y + {\alpha(\omega)} \left( \frac{y_1 +\lambda_y y}{x_1 + \lambda_x x}\lambda_x x -
  \lambda_y y\right).
   \end{equations} 
Note that, denoting 
\begin{equations}\label{AA1}
  &x_1^* = x_1 + { \beta(\omega)} \left( \frac{x +x_1}{y + y_1} y_1 - x_1\right) ,\\
   & y_1^* = y_1 + {\alpha(\omega)} \left( \frac{y +y_1}{x + x_1} x_1 -
   y_1\right).
   \end{equations}
exchanging variables into the integral one has the identity
 \[
\left\langle \int_{\real_+^4}  (x^* -x )f(x,y,t)f(x_1,y_1,t) \, d\pi \,\right\rangle = 
 \]     
 \[
\frac 12 \left\langle \int_{\real_+^4}  (x^*+x_1^* -x-x_1 )f(x,y,t)f(x_1,y_1,t) \, d\pi\,\right\rangle = 0.
 \]
This property is not satisfied by the mixed operators $P(f.g)$ and $\bar P(g,f)$. However it can be easily verified that
 \[
\int_{\real_+^2}x P(f,g)(x,y) \,dx  dy + \int_{\real_+^2}x \bar P(g,f)(x,y) \,dx  dy = 0.
 \]
The same properties hold if we substitute the good $x$ with the good $y$. This implies that the 
conservation of the mean values, as given by \fer{conXY} still hold for the nonlinear system . Since the exchanges
of goods of type \fer{trB} are nonlinear, the study of the properties of
the solution to the Boltzmann system appears  difficult. The Boltmann system \fer{we-f}, \fer{we-g} can however be studied from a numerical point of view.
 
\section{Numerical experiments}\label{sec-num}
This section contains a numerical description of the solutions  to the nonlinear Boltzmann system \fer{we-f} and \fer{we-g}. For the numerical approximation of the Boltzmann equations we apply a Monte Carlo method, as described in Chapter 4 of \cite{PT13}. If not otherwise stated the kinetic simulation has been performed with $N=10^4$ agents.

The numerical experiments will help to clarify the effect of the strategy of speculators in the final distribution of the wealth density among the two classes of agents. It is evident from the experiments that, thanks to the conservations of the mean
quantities of the goods, the densities $f(x,y,t)$ of dealers and $g(x,y,t)$ of speculators will converge to stationary distributions \cite{PT13}. As usual in kinetic theory,
these stationary solutions will be reached in an exponentially fast
time.
We will evaluate the stationary solutions  for different values of the parameters $\lambda_x,
\lambda_y$ and different values of the parameters $\alpha$ and $\beta$.
In this way we will recognize first the effect of the strategy of speculators and, second,  the role of the preferences parameters $\alpha$ and $\beta$ to reach the final distribution of wealth.

The main test represents a two-phases experiment. In the first phase, the convergence of the price to its equilibrium value is shown by considering only the population of dealers. Then, when the price is close to its equilibrium value, the speculators enter into the game to modify the price value. At difference with the linear model described in Section \ref{sec-lin}, where in absence of the population of speculators the price of the goods in possess of the population of dealers takes a constant value, the nonlinear model exhibits oscillations in the price evolution, which reduce exponentially in time.

The test is intended to simulate the situation in which a small number of speculators are entering into the market to obtain a marked advantage in their wealth by a saving politics. Since the price of the goods is adapting exponentially fast in time, the experiment also justifies the fact that dealers have no time to adapt to the new situation by changing their preferences.

 \begin{figure}
\centering
\includegraphics[width=0.25\textwidth]{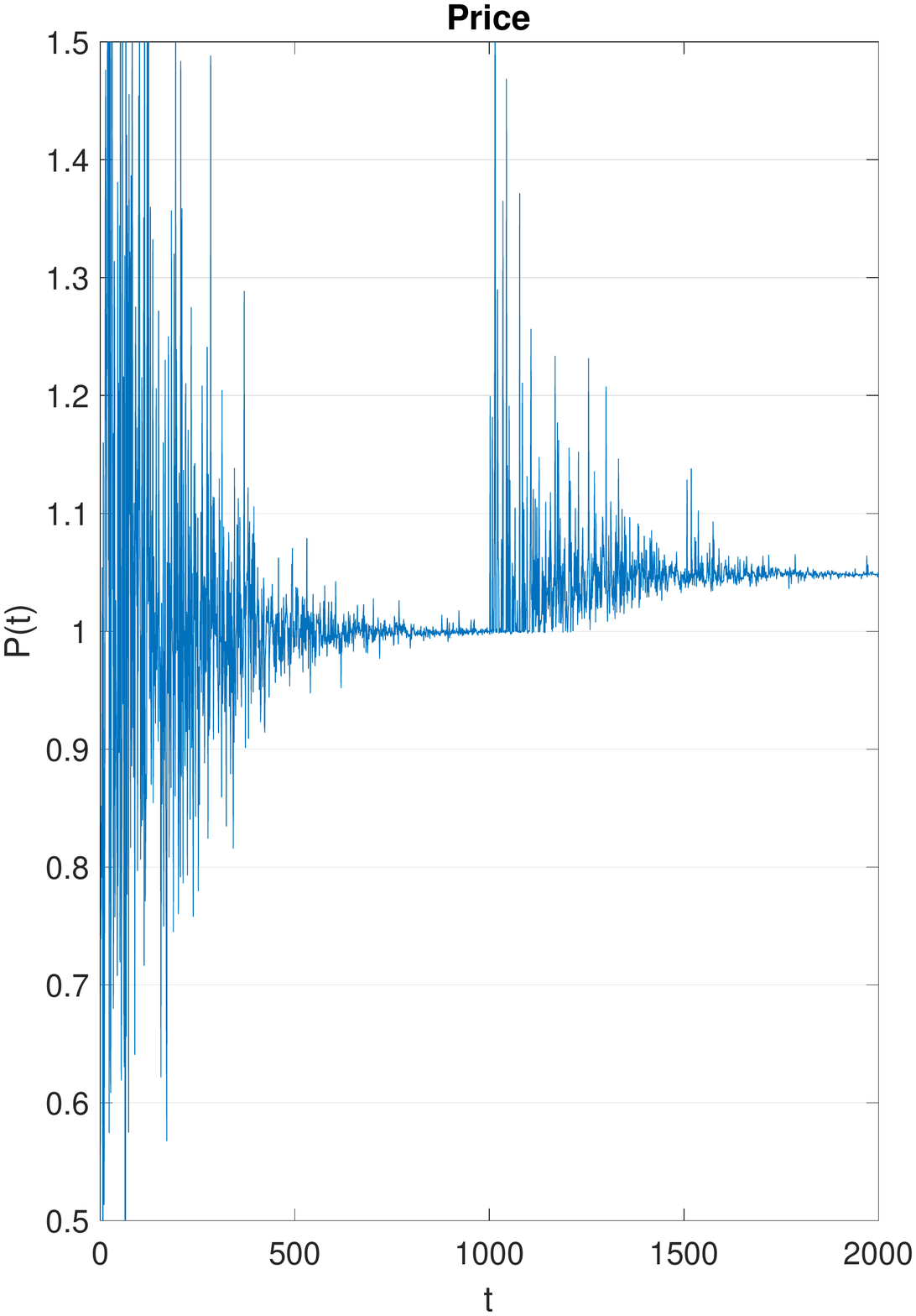}%
\includegraphics[width=0.25\textwidth]{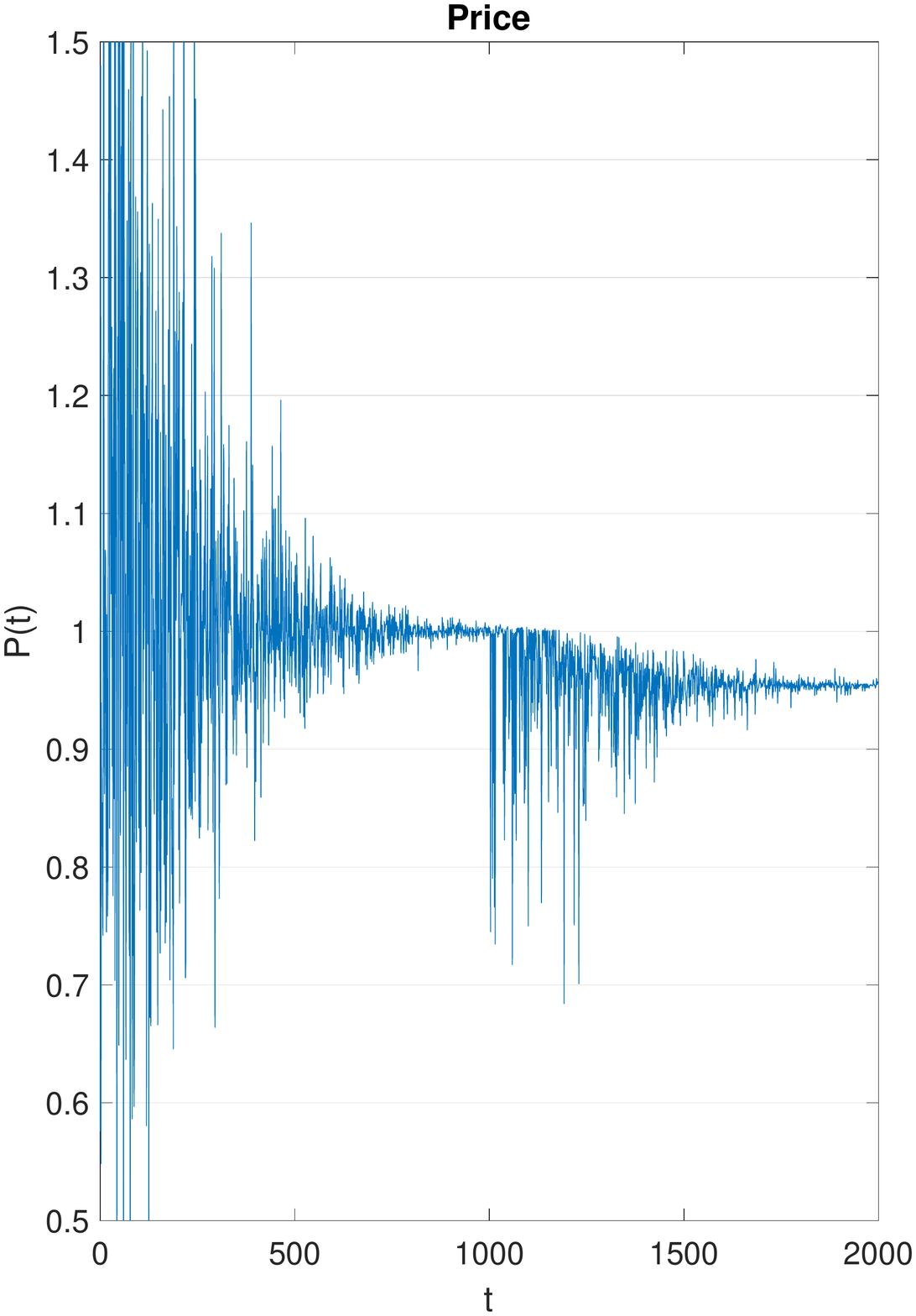}
\caption{$\alpha = 0.5; \beta = 0.5; M_x =3; m_x = 10; M_y= 3; m_y= 2; \lambda_x = 0.8; \lambda_y= 0.2$ (left);
$\alpha = 0.5; \beta = 0.5; M_x =3; m_x = 10; M_y= 3; m_y= 2; \lambda_x = 0.5; \lambda_y= 0.5$ (right).}
\label{fig:price goes up and down}
\end{figure}

The test is performed for different values of the relevant parameters $\alpha$, $\beta$, $\lambda_x$ and $\lambda_y$, to clarify their effect on the price evolution. 

Figure \ref{fig:price goes up and down} shows the variation of the (relative) price of the second good induced by the strategy of speculators. When the  saving parameters $\lambda_x$ and $\lambda_y$ are such that $\lambda_x > \lambda_y$, the price of the good $Y$ is shown to increase (left). On the contrary, by taking the saving parameters  equal, and by leaving the other quantities unchanged, the price of the good $Y$ is shown to increase (left). Note that in both cases the value of the price consequent to the action of the speculators decays exponentially fast towards the limit value.

\begin{figure}
\centering
\includegraphics[width=0.25\textwidth]{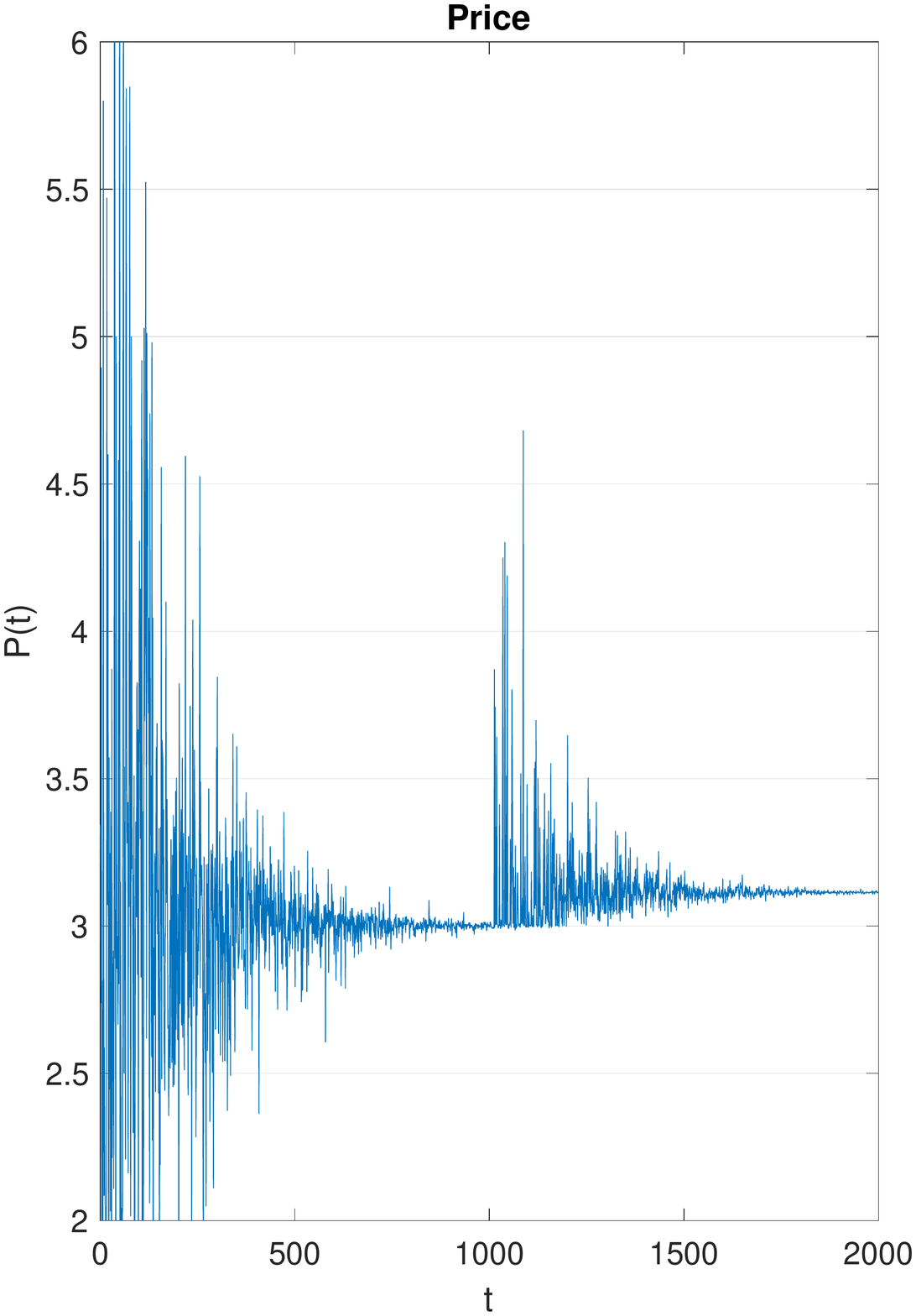}%
\includegraphics[width=0.25\textwidth]{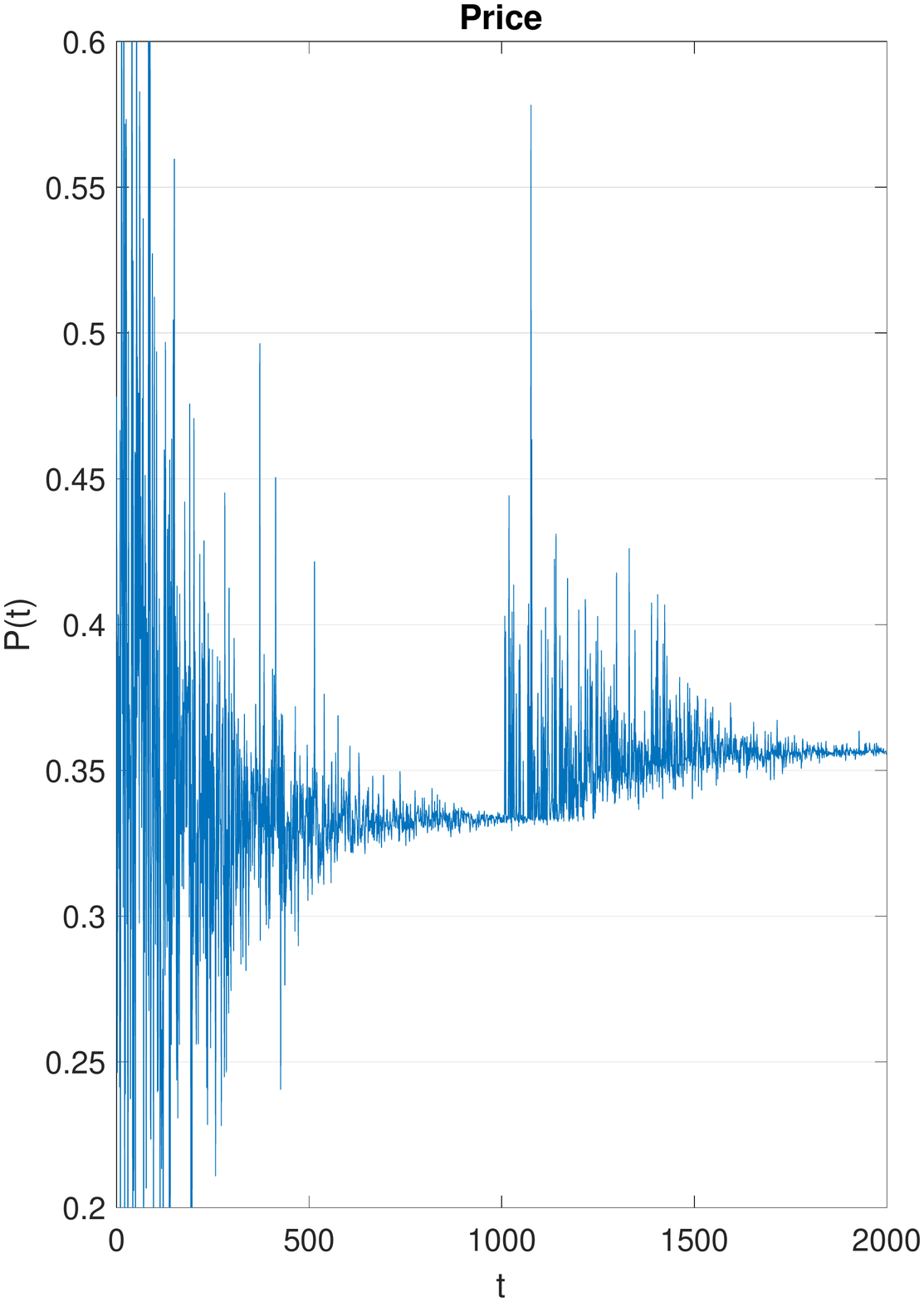}
\caption{$\alpha = 0.25; \beta = 0.75; M_x =3; m_x = 10; M_y= 3; m_y= 2; \lambda_x = 0.8; \lambda_y= 0.2$ (left);
$\alpha = 0.75; \beta = 0.25; M_x =3; m_x = 10; M_y= 3; m_y= 2; \lambda_x = 0.5; \lambda_y= 0.5$ (right).}
\label{fig:preferences}
\end{figure}

Figure \ref{fig:preferences} shows the variation of the (relative) price of the second good induced by the strategy of speculators in presence of a radical change of the preference parameters $\alpha$ and $\beta$. All the remaining parameters are left equal. While both experiments lead to a positive variation of the (relative) price of the second good, the final price in the two experiments is completely different. While in the case to the left, denoted by a marked preference for the good $Y$, the price of this good stabilizes around a value between $3$ and $3.5$,  in the case to the right, characterized by a marked preference for the good $X$, the price of the good $y$ stabilizes around a value $0.35$, namely a factor ten below the prive of the experiment on the left.  

\begin{figure}
\centering
\includegraphics[width=0.25\textwidth]{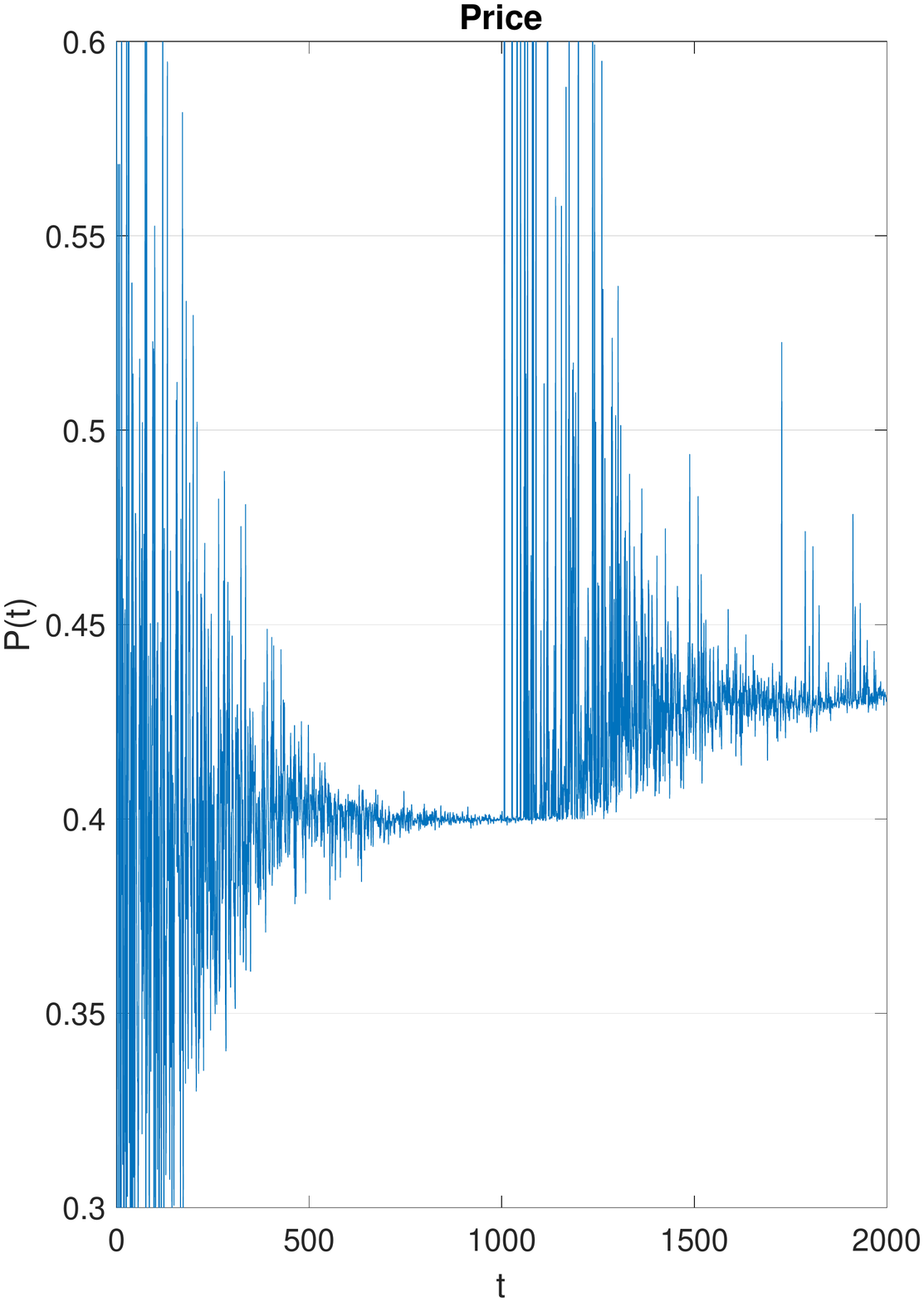}%
\includegraphics[width=0.25\textwidth]{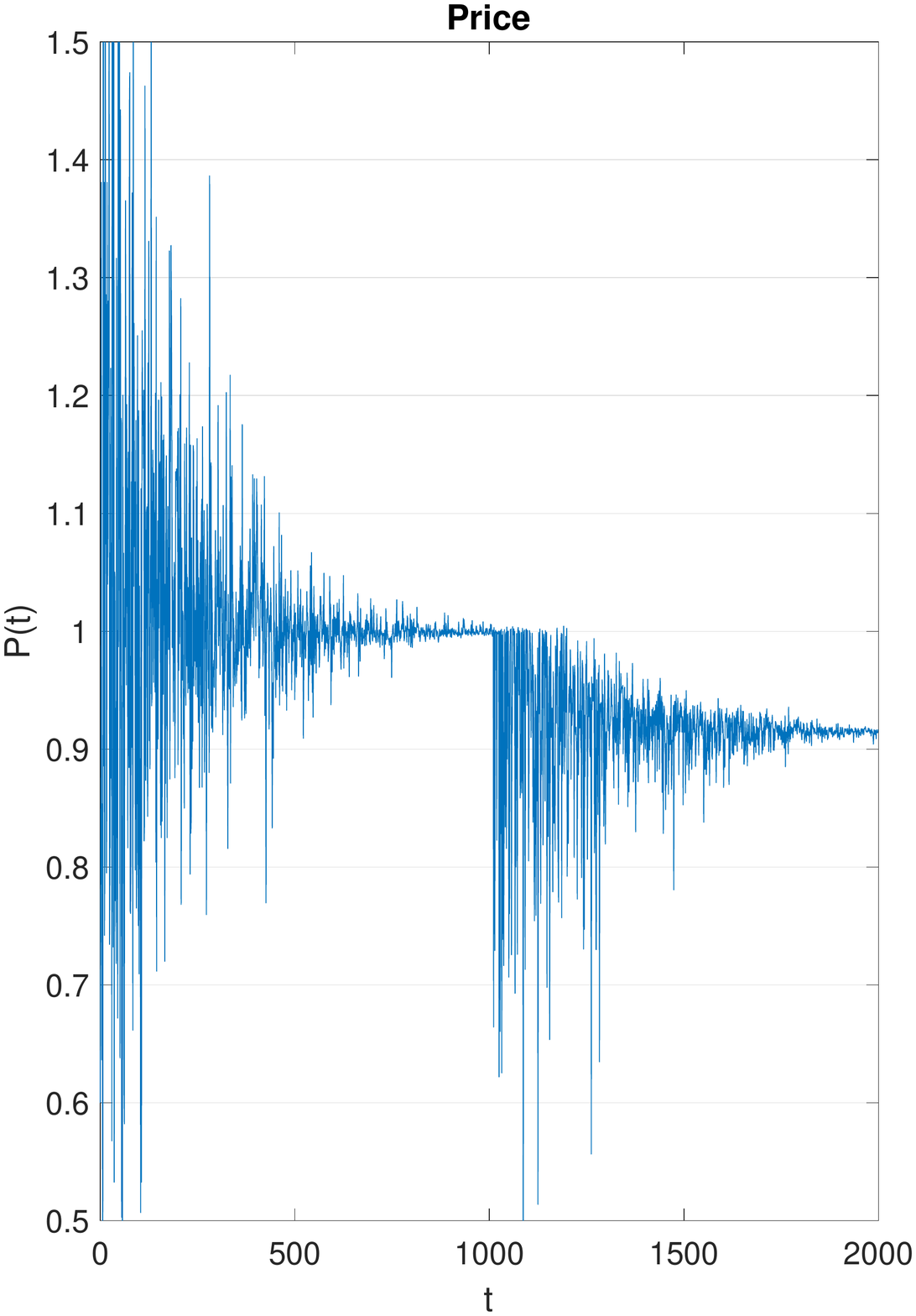}
\caption{$\alpha = 0.5; \beta = 0.5; M_x =3; m_x = 20; M_y= 7,5; m_y= 5; \lambda_x = 0.8; \lambda_y= 0.2$ (left);
$\alpha = 0.5; \beta = 0.5; M_x =3; m_x = 7.5; M_y= 20; m_y= 5; \lambda_x = 0.2; \lambda_y= 0.8$ (right).}
\label{fig:last}
\end{figure}

Last, Figure \ref{fig:last} shows the evolution of price in the case of different choices of the saving parameters, coupled with different mean quantities of goods. As in the case of Figure \ref{fig:price goes up and down} the relevant parameters which allow to increase (or decrease the final relative price of good $Y$ seem to be the saving parameters $\lambda_x$ and $\lambda_y$. 

\section{Final remarks}
In this paper we introduced two systems of  kinetic equations of Boltzmann type suitable to describe the evolution of the probability distribution of two goods among two populations of agents, dealers and speculators, that apply different strategies in the exchanges. The leading idea was to describe the trading of these goods
by means of some fundamental rules in prize theory, in particular by
using Cobb-Douglas utility functions for the binary exchange. Also, to take into
account the intrinsic risks of the market, we introduced randomness
in the exchange, without affecting the microscopic conservations, that is the conservation of the total number of each good in the market.

Both the analytic study of the linear system \fer{we}, and the numerical simulation of the nonlinear system \fer{we-f} and \fer{we-g} allow to conclude that the saving politics of the speculators is able to modify the final price of the goods, to achieve eventually a net gain.


\section*{Acknowledgement} This work has been written within the
activities of GNFM group  of INdAM (National Institute of
High Mathematics), and partially supported by  MIUR project ``Optimal mass
transportation, geometrical and functional inequalities with applications''.


\end{document}